\documentclass[journal]{vgtc}                % final (journal style)
\usepackage{mathptmx}
\usepackage{graphicx}
\usepackage{times}

\usepackage{amsfonts}
\usepackage{amsmath}
\usepackage{amssymb}
\usepackage{amsthm}

\usepackage{cite}

\usepackage{booktabs}

\graphicspath{ {Figures/} }

\usepackage[bookmarks,backref=true,linkcolor=black]{hyperref} %,colorlinks
\hypersetup{
  pdfauthor = {},
  pdftitle = {},
  pdfsubject = {},
  pdfkeywords = {},
  colorlinks=true,
  linkcolor= black,
  citecolor= black,
  pageanchor=true,
  urlcolor = black,
  plainpages = false,
  linktocpage
}

\newcommand{\mathsc}[1]{{\normalfont\textsc{#1}}}

\onlineid{122}

\vgtccategory{Theory}

\vgtcinsertpkg

\title{What May Visualization Processes Optimize?}

\author{Min Chen, \textit{Member, IEEE}, and Amos Golan}
\authorfooter{
\item	Min Chen is with University of Oxford,
	E-mail: min.chen@oerc.ox.ac.uk.
\item	Amos Golan is with American University,
	E-mail: agolan@american.edu.
}

\shortauthortitle{Chen and Golan: What May Visualization Processes Optimize?}
\abstract{In this paper, we present an abstract model of visualization and inference processes and describe an information-theoretic measure for optimizing such processes. In order to obtain such an abstraction, we first examined six classes of workflows in data analysis and visualization, and identified four levels of typical visualization components, namely disseminative, observational, analytical and model-developmental visualization. We noticed a common phenomenon at different levels of visualization, that is, the transformation of data spaces (referred to as alphabets) usually corresponds to the reduction of maximal entropy along a workflow.  Based on this observation, we establish an information-theoretic measure of cost-benefit ratio that may be used as a cost function for optimizing a data visualization process. To demonstrate the validity of this measure, we examined a number of successful visualization processes in the literature, and showed that the information-theoretic measure can mathematically explain the advantages of such processes over possible alternatives.
} % end of abstract


\begin{document}


\firstsection{Introduction}

\maketitle

%% \section{Introduction} %for journal use above \firstsection{..} instead

% is a branch of probability theory \cite{Kullback:1968:book}.
Over the past 25 years, the field of \emph{visualization} has developed to encompass three major subfields, namely \emph{scientific visualization}, \emph{information visualization} and \emph{visual analytics} as well as many domain-specific areas, such as geo-information visualization, biological data visualization, software visualization, and others.
A number of pipelines have been proposed for visualization in general (e.g., \cite{Upson:1989:CGA,vanWijk:2005:Vis,Moreland:2013:TVCG}) and for visual analytics in particular \cite{Keim:2008:LNCS,Green:2008:VAST}.
In practice, a visualization workflow normally includes machine-centric components (e.g., statistical analysis, rule-based or policy-based models, and supervised or unsupervised models) as well as human-centric components (e.g., visualization, human-computer interaction, and human-human communication).
The integration of these two types of components become more and more common since \emph{visual analytics} \cite{Thomas:2005:book,Wong:2004:CGA} has become a de facto standard approach for handling large volumes of complex data.

Given a visualization workflow in a specific context, it is inevitable that one would like to improve its cost-benefit ratio, from time to time, in relation to many factors such as accuracy, speed, computational and human resources, creditability, logistics, changes in the environment, data or tasks concerned, and so forth. Such improvement can typically be made through introducing new technologies, restructuring the existing workflow, or re-balancing the tradeoff between different factors.
While it is absolutely essential to optimize each visualization workflow in a heuristic and case-by-case manner \cite{Munzner:2009:TVCG}, it is also desirable to study the process optimization theoretically and mathematically through abstract reasoning.
In many ways, this is similar to the process optimization in tele- and data communication, where each subsystem is optimized through careful design and customization but the gain in cost-benefit is mostly underpinned by information theory \cite{Shannon:1948:BSTJ,Cover:2006:book}.
In this paper, we study, in abstraction, the process optimization in visualization from an information-theoretic perspective.

Visualization is a form of information processing.
Like other forms of information processing (e.g., statistical inferences), visualization enables transformation of information from one representation to another.
The objective of such a transformation is typically to infer a finding, judgment or decision from the observed \emph{data}, which may be incomplete and noisy.
The input to the transformation may also include ``soft'' \emph{information} and \emph{knowledge}, such as known theories, intuition, belief, value judgment, and so on.
Another form of input, which is often referred to as \emph{priors}, may come from knowledge about the system where the data are captured, facts about the system or related systems, previous observations, experimentations, analytical conclusions, etc.
Here we use the terms \emph{data}, \emph{information} and \emph{knowledge} according to the commonly-used definitions in computational spaces \cite{Chen:2009:CGA}.

All inferential processes are designed for processing a finite amount of information.
In practice, they all encounter some difficulties, such as
the lack of adequate technique for extracting meaningful information from a vast amount of data;
incomplete, incorrect or noisy data;
biases encoded in computer algorithms or biases of human analysts;
lack of computational resources or human resources;
urgency in making a decision; and so on.
All inferential problems are inherently under-determined problems \cite{Golan:2008:book,Golan:2012:LNCS}.

The traditional machine-centric solutions to the inferential problem address these difficulties by imposing certain assumptions and structures on the model of the system where the data are captured.
If these assumptions were correctly specified and these structures were perfectly observed, computed inference based on certain statistics (e.g., moments) would provide us with perfect answers.
In practice, it is seldom possible to transform our theory, axioms, intuition and other soft information into such statistics.
Hence optimization of a visualization process is not just about the best statistical method, the best analytical algorithm, or the best machine learning technique.
It is also about the best human-centric mechanisms for enabling uses of ``soft'' information and knowledge.

In this paper, we propose to measure the cost-benefit of a visualization-assisted inference process within an information-theoretic framework.
The work is built on a wealth of literature on visualization and visualization pipelines (e.g., \cite{Upson:1989:CGA,vanWijk:2005:Vis,Moreland:2013:TVCG,Keim:2008:LNCS,Green:2008:VAST}) and that on information theoretic measures and inference in the statistics and econometrics \cite{Golan:1996:book,Golan:2000:JBES,Hansen:1982:E}.
It is a major extension of the information-theoretic framework for visualization proposed by Chen and J\"{a}nicke \cite{Chen:2010:TVCG}, and a major extension of statistical inference and information processing in general (e.g., \cite{Golan:2008:book}).
Our contributions are:

\begin{itemize}
\vspace{-2mm}
\item We propose a new categorization of visualization workflows and identify four levels of visualization commonly featured in different data analysis and visualization processes (Section \ref{sec:Workflow}).
\vspace{-2mm}
\item We present an information-theoretic abstraction of visualization processes as transformation of alphabets along a workflow for data analysis and visualization, and identify a common trend of reduction of Shannon entropy (i.e., uncertainty) in such workflows  (Section \ref{sec:Abstraction}).
\vspace{-2mm}
\item We propose an information-theoretic measure of cost-benefit, which can be applied to the whole workflow as well as individual processing steps (Section \ref{sec:Abstraction}).
\vspace{-2mm}
\item We demonstrate that this cost-benefit measure can explain the information-theoretic advantages of successful visualization workflows in the literature, suggesting that it can be used for optimizing a visualization-assisted inference process through a combination of quantitative and qualitative analysis (Section \ref{sec:Demonstration}). 
\end{itemize}

% ===========
\section{Related Work}
\label{sec:RelatedWork}

In 2003, Grinstein \emph{et al.} \cite{Grinstein:2003:Vis} posed an intriguing question about usability vs. utility when they considered visualization as an interface technology that draws from both machine- and human-centric capabilities.
This is a question about optimization.
 
\vspace{1mm}
\noindent \textbf{Pipelines and Workflows.}
In the field of visualization, many have considered pipelines or workflows that feature components such as analysis, visualization and interaction.
Upson \emph{et al.} provided one of the earliest abstraction of a pipeline with four main components, data source, filtering and mapping, rendering and output \cite{Upson:1989:CGA}.
Wood et al. proposed an extension for collaborative visualization in the form of parallel pipelines \cite{Wood:1997:Vis}.
van Wijk outlined a two-loop pipeline, bringing interaction and cognition into a visualization process \cite{vanWijk:2005:Vis}.
Green \emph{et al.} proposed a revision of this pipeline \cite{Green:2008:VAST}.
Keim \emph{et al.} proposed a pipeline featuring two interacting parallel components for data mining models and visual data exploration respectively \cite{Keim:2008:LNCS}.
J\"anicke \emph{et al.} examined several pipelines for comparative visualization, and discussed quality metrics for evaluating reconstructibility of visualization \cite{Jaenicke:2011:CGF}.
Bertini \emph{et al.} proposed an automated visualization pipeline driven by quality metrics \cite{Bertini:2011:TVCG}.
Recently Moreland surveyed visualization pipelines mainly in the context of scientific visualization \cite{Moreland:2013:TVCG}.
There are many other variations of visualization pipelines in the literature, such as \cite{Card:1997:InfoVis,Chi:2000:InfoVis,Groth:2006:TVCG,JankunKelly:2007:TVCG,Chen:2013:S}.
All these discussions on visualization pipelines pointed out one common fact, i.e., visualization processes can be broken down to steps, which may be referred to as transformations or mappings.
This work considers this ubiquitous feature of visualization in abstraction.     

\vspace{1mm}
\noindent \textbf{Design Methods and Processes.}
Abram and Treinish proposed to implement visualization processes on data-flow architectures \cite{Abram:1995:Vis}.
Chi described visualization processes using a state reference model, involving data, visualization, and visual mapping transformation \cite{Chi:2000:InfoVis}.
Jansen and Dragicevic proposed an interaction model in the context of visualization pipelines \cite{Jansen:2013:TVCG}.
Munzner proposed a nested model for designing and developing visualization pipelines \cite{Munzner:2009:TVCG}.
Wang \emph{et al.} proposed a two-stage framework for designing visual analytics systems \cite{Wang:2011:VAST}.
Ahmed \emph{et al.} proposed to use purpose-driven games for evaluating visualization systems \cite{Ahmed:2012:TVCG}.
Scholtz outlined a set of guidelines for assessing visual analytics environments \cite{Scholtz:2011:JIV},
and Scholtz \emph{et al.} further developed them into an evaluation methodology \cite{Scholtz:2013:JIV}.
The theoretic abstraction presented in this paper is built on these works, and complement them by offering a mathematical rationalization for good practices in designing and assessing visualization systems.   

\vspace{1mm}
\noindent \textbf{Theories of Visualization and their Applications.}
In developing theories of visualization, much effort has been made in formulating categorizations and taxonomies (e.g., \cite{Bertin:1983:book,Ware:2013:book,Tory:2004:InfoVis}).
Some 25 different proposals are listed in \cite{Chen:2010:TVCG,Chen:2014:CGF}.
In addition, a number of conceptual models have been proposed, including
object-oriented model by Silver \cite{Silver:1995:CGA},
feature extraction and representation by van Walsum \emph{et al.} \cite{vanWalsum:1996:TVCG},
visualization exploration by Jankun-Kelly \emph{et al.} \cite{JankunKelly:2007:TVCG},
distributed cognition model by Liu \emph{et al.} \cite{Liu:2008:TVCG},
predictive data-centered theory by Purchase \emph{et al.} \cite{Purchase:2008:LNCS},
Visualization Transform Design Model by Purchase \emph{et al.} \cite{Purchase:2008:LNCS},
cognition model for visual analytics by Green \emph{et al.} \cite{Green:2009:JIV},
sensemaking and model steering by Endert \emph{et al.} \cite{Endert:2012:TVCG},
modelling visualization using semiotics and category theory by Vickers \emph{et al.} \cite{Vicker:2013:TVCG},
composition of visualization tasks by Brehmer and Munzner \cite{Brehmer:2013:TVCG}, and
visual embedding by Demiralp \emph{et al.} \cite{Demiralp:2014:CGA}. 
Recently, Sacha \emph{et al.} proposed a knowledge generation model \cite{Sacha:2014:TVCG}, introducing a visual analytics model with exploration and verification loops.
The deliberations in these works represent qualitative abstraction of visualization processes.   

Meanwhile, the development of mathematical frameworks is gathering its pace in recent years.
One of these is the information theoretic framework, which was initially suggested by Ward \cite{Purchase:2008:LNCS}, then generalized and detailed by Chen and J\"anicke \cite{Chen:2010:TVCG}, and further enriched by Xu \emph{et al.} \cite{Xu:2010:TVCG} and Wang and Shen \cite{Wang:2011:E} in the context of scientific visualization.
Another is the algebraic framework proposed by Kindlmann and Scheidegger \cite{Kindlmann:2014:TVCG}, who justifiably placed their focus on visual mappings, which are inherently the most important transformations from a visualization perspective.
While an algebraic formulation typically describes mappings between set members (e.g., from a pair of datasets to a pair of visual representations in \cite{Kindlmann:2014:TVCG}), an information-theoretic formulation describes mappings between sets together with the probabilistic distributions of their members.

This holistic nature of information-theoretic reasoning has enabled many applications in visualization, including
% scene and shape complexity analysis by Feixas \emph{et al.} \cite{Feixas:2001:CGF} and Rigau \emph{et al.} \cite{Rigau:2005:SMA},
light source placement by Gumhold \cite{Gumhold:2002:Vis},
view selection in mesh rendering by V\'{a}zquez \emph{et al.} \cite{Vazquez:2004:CGF} and Feixas \emph{et al.} \cite{Feixas:2009:AP},
% attribute selection by Ng and Martin \cite{Ng:2004:IV},
view selection in volume rendering by Bordoloi and Shen \cite{Bordoloi:2005:Vis}, and Takahashi and Takeshima \cite{Takahashi:2005:Vis},
focus of attention in volume rendering by Viola \emph{et al.} \cite{Viola:2006:TVCG},
multi-resolution volume visualization by Wang and Shen \cite{Wang:2006:TVCG},
feature highlighting in unsteady multi-field visualization by J\"anicke and Scheuermann \cite{Jaenicke:2007:TVCG,Jaenicke:2010:CGA},
feature highlighting in time-varying volume visualization by Wang \emph{et al.} \cite{Wang:2008:TVCG},
transfer function design by Bruckner and M\"{o}ller \cite{Bruckner:2010:CGF},
	and by Ruiz \emph{et al.} \cite{Ruiz:2011:TVCG,Bramon:2013:JBHI},
multimodal data fusion by Bramon \emph{et al.} \cite{Bramon:2012:TVCG},
evaluating isosurfaces \cite{Wei:2013:CGF},
measuring of observation capacity \cite{Bramon:2013:CGF},
measuring information content in multivariate data \cite{Biswas:2013:TVCG}, and
confirming the mathematical feasibility of visual multiplexing \cite{Chen:2014:CGF}.

% ===========
\section{Workflows in Visualization}
\label{sec:Workflow}

\begin{figure}[t!]
\centering
\includegraphics[width=\columnwidth]{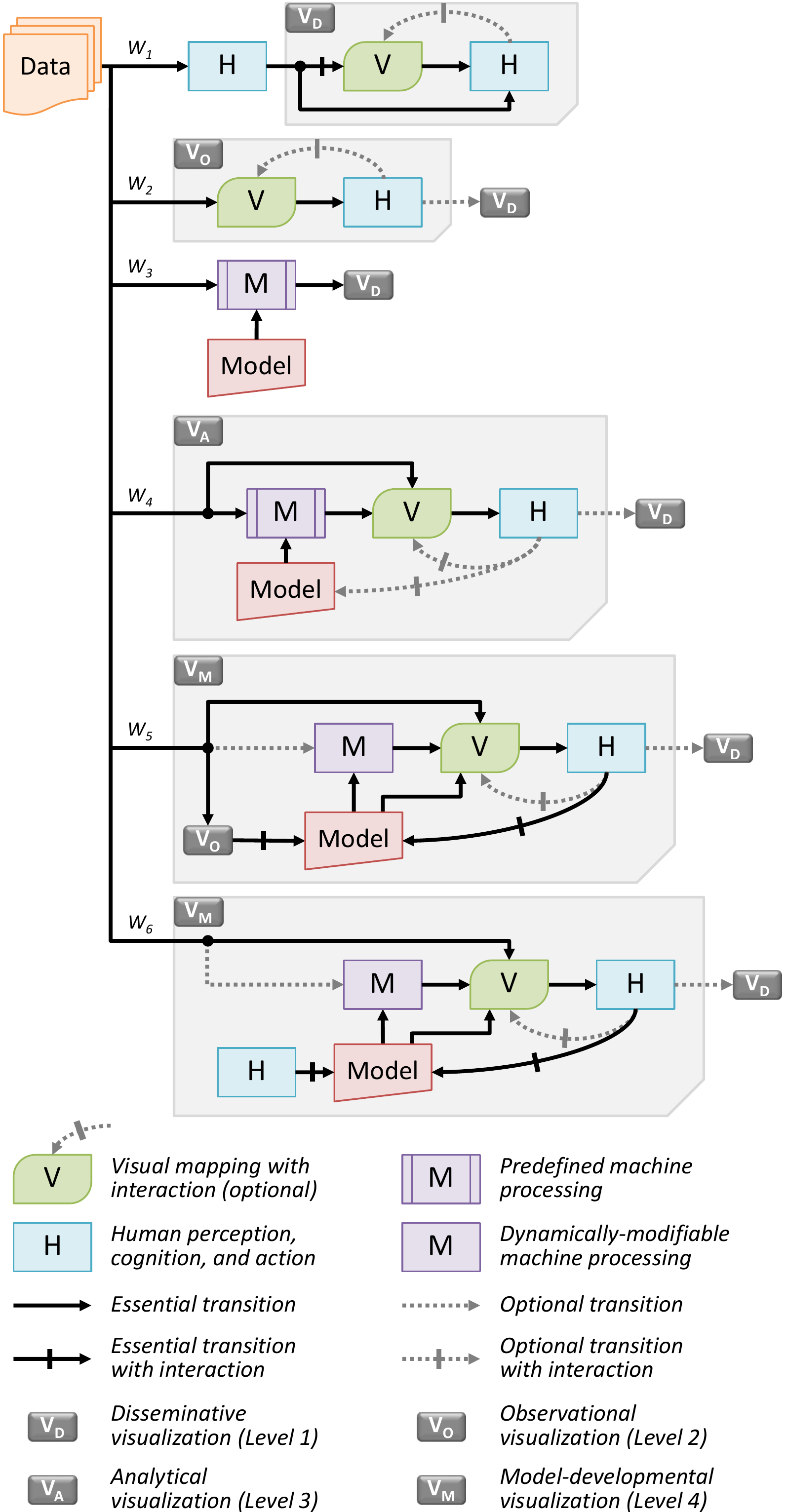} \\
\caption{Six typical workflows in data analysis and visualization. The subgraphs, $\mathbf{V_D}$, $\mathbf{V_O}$, $\mathbf{V_A}$, and $\mathbf{V_M}$ represent four levels of visualization.}
\label{fig:Workflows}
\vspace{-4mm}
\end{figure}
% ----------
\subsection{Six Classes of Workflows}
Consider a broad range of workflows in visualization, including those historically referred to as analysis, inference, simulation or visual analytics as well as those emerged recently, as long as they feature a component of \emph{visualization}, i.e., mapping some data to alternative visual representations.
As a process of abstraction, we group these workflows into six classes as illustrated in Fig. \ref{fig:Workflows}.
They feature the following types of components:

\begin{itemize}
\vspace{-2mm}
\item \emph{Machine Processing} (M) ---
These are computational processes executed by computers including, for instance, computation of statistical indicators (e.g., mean, correlation index, etc.), data analysis (e.g., classification, anomaly detection, association analysis, etc.), simulation, prediction, recommendation and so on.
Each computational process is defined by a program that may encode a theoretic or heuristic model, which we refer to generally as a \emph{Model}.
\vspace{-2mm}
\item \emph{Human Processing} (H) ---
These are human cognitive processes and related activities including, for instance, viewing, reasoning, memorizing, discussing, decision making and so on.
\vspace{-2mm}
\item \emph{Visual Mapping} (V) --- These are processes where data are transformed to alternative visual representations to be viewed by humans. We purposely treat these processes separately from M and H, and assume that visual representations can be generated by many means from hand-drawn plots and illustrations to automated generation of visualization.
\vspace{-2mm}
\item \emph{Interaction} (I) --- These are actions taken by humans to influence an M or V process. They include typical interactions in visualization \cite{Yi:2007:TVCG}, such as parameter adjustment, and model creation and refinement.
In Fig. \ref{fig:Workflows}, they are not explicitly shown as a processing block, as the main cognitive processing for interaction is assumed to take place in H. Instead, they are indicated by a solid bar on a connection.
\end{itemize}

Workflow class $W_1$ encompasses perhaps some of the most common process in data analysis and visualization.
In this process, one or more human analysts (H) process the input data with or without the aid of computation, gain some understanding, create some visualization (V) and convey the understanding to others (H).
Many visualization images in the media and visual illustration in user manuals fall into this class.
The goal of visualization is to pass on known information and knowledge to others, and the dissemination process is almost always accompanied by written or verbal commentaries describing the understanding and/or opinions of analysts. 
We refer to this form of visualization as \emph{Disseminative Visualization}, and represent the visualization part of the workflow as a macro block $\mathbf{V_D}$.

The second class, $W_2$, encompasses many operational processes, where human analysts need to use visualization to observe data routinely.
For examples, stock brokers frequently glance at various time series plots, drivers glance at their GPS-navigation devices regularly, neurologists examine visual representations of various scans (e.g., electroencephalography, computed tomography, diffusion tensor imaging, etc.) of their patients, and computational scientists visualize simulation results after each run.
The goal of visualization is to enable intuitive and speedily observation of features, phenomena and events captured in the data, and to provide external memorization of what have been observed.
We refer to this form of visualization as \emph{Observational Visualization}, and represent this as a macro block $\mathbf{V_O}$.
Although the two macro blocks $\mathbf{V_O}$ and $\mathbf{V_D}$ appear to be similar except an extra forward transition in $\mathbf{V_D}$, their fundamental difference is that in $\mathbf{V_D}$ analysts have already gained the understanding to be conveyed before the visualization is generated, while in $\mathbf{V_O}$ visualization is generated in order to gain a new understanding.
Of course, $\mathbf{V_O}$ can be followed by $\mathbf{V_D}$ to disseminate such a new understanding.

Workflow $W_3$ depicts a class of processes where automated data analysis plays a dominant role, and humans are only the destination of dissemination.
In many ways, $W_3$ is almost identical to $W_1$, except that in $W_3$ the understanding and/or opinions conveyed to humans through $\mathbf{V_D}$ are from machine processing.
Such a workflow has its place in data analysis and visualization, when the machine is always or almost always correct about what is being conveyed.
When such a high level of correctness is not assured, it is necessary to increase humans' involvement in these processes.

This leads to workflow class $W_4$, where human analysts are able to observe input data in conjunction with the machine's ``understanding''.
In many ways, this workflow is similar to the parallel pipeline proposed by Keim \emph{et al.} \cite{Keim:2008:LNCS}.
It allows analysts to receive computational results from machine processing, while evaluating the correctness of the results and identify possible false positives and negatives.
For example, in much investigative analysis for examining and understanding complex relationships among data objects, the amount of input data often makes direct observation time-consuming.
The machine-processing hence enables the analysts to prioritize their effort and structure their reasoning and decision-making process.
At the same time, analysts are able to explore the data and adjust the model depending on the analysts' judgment about the quality of the computed results.
We refer to this form of visualization as \emph{Analytical Visualization}, and represent this as a macro block $\mathbf{V_A}$.

When the correctness or accuracy of a model is the main concern, the focus of visualization is shifted to assisting analysts in improving an existing model or creating a new model.
Both workflow classes $W_5$ and $W_6$ represent such a focus.
In $W_5$, analysts first observe some input data, and then identify an existing model or formulate a new one for processing the data.
Tasks for such processing may include, but not limited to, computing statistical indicators; detecting features, objects, and events; identifying patterns, associations, and rules; and making predictions and recommendations. 
In many cases, $W_5$ may represent a long-term process for developing a theory and its applications, such as physical laws and their applications in computer simulation.
$W_6$ thus represents a class of commonly-occurred workflows where analysts deploy known theories to specify a model without the initial observational visualization for establishing these theories.  
In practice, to create, test and optimize a model, analysts often make use of $W_5$ and $W_6$ for different parts of a model.
For example, in developing a simulation model, major computation steps are defined according to known quantitative laws, while initial and boundary conditions are defined based on observations and experiments. 
We thereby refer to these two forms of visualization collectively as \emph{Model-developmental Visualization}, and represent them as the same macro block $\mathbf{V_M}$.
Note that we have avoided the phrase ``modelling visualization'' here as it could be misread as an action ``to model visualization''.
One day, there might be a new adjective, e.g., in the form of \emph{modelative} or \emph{modelary}.

% ----------
\subsection{Four Levels of Visualization}
The four macro blocks, namely $\mathbf{V_D}$,  $\mathbf{V_O}$,  $\mathbf{V_A}$, and  $\mathbf{V_M}$ can be seen as \emph{four levels of visualization}.
The different levels, which are summarized below, reflect the complexity of visualization tasks from the perspective of analysts.  

\begin{itemize}
\vspace{-2mm}
\item \emph{Level 1: Disseminative Visualization} ($\mathbf{V_D}$) ---
Visualization is a presentational aid for disseminating information or insight to others.
The analyst who created the visualization does not have a question about the data, except for informing others: ``This is $A$!'' where $A$ may be a fact, a piece of information, an understanding, etc.  
At this level, the complexity for the analyst to obtain an answer about the data is O(1).
Here we make use the big O notation in algorithm and complexity analysis.   
\vspace{-2mm}
\item \emph{Level 2: Observational Visualization} ($\mathbf{V_O}$) ---
Visualization is an operational aid that enables intuitive and/or speedily observation of captured data.
It is often a part of routine operations of an analyst, and the questions to be answered may typically be in the forms of ``What has happened?'' "When and where $A$, $B$, $C$, etc., happened?'
At this level, the observation is usually sequential, and thus the complexity is generally O($n$), where $n$ is the number of data objects.
Broadly speaking, a \emph{data object} is a data record.
We will give a more precise definition of it in Section \ref{sec:Abstraction}.
\vspace{-2mm}
\item \emph{Level 3: Analytical Visualization} ($\mathbf{V_A}$) ---
Visualization is an investigative aid for examining and understanding complex relationships (e.g., correlation, association, causality, contradiction).
The questions to be answered are typically in the forms of ``What does $A$ relate to?'' and ``Why?''    
Given $n$ data objects, the number of possible $k$-relationships among these data objects is at the level of O($n^k$) ($k \geq 2$).
For a small $n$, it may be feasible to examine all $k$-relationships using observational visualization.
When $n$ increases, it becomes necessary to use analytical models to prioritize the analyst's investigative effort.
Most visual analytics processes reported in the recent literature operate at this level.
\vspace{-2mm}
\item \emph{Level 4: Model-developmental Visualization} ($\mathbf{V_M}$) ---
Visualization is a developmental aid for improving existing models, methods, algorithms and systems, as well as for creating new ones.
The questions to be answered are typically in the forms of ``How does $A$ lead to $B$?'' and ``What are the exact steps from $A$ to $B$?''
If a model has $n$ parameters and each parameter may take $k$ values, there are a total of $k^n$ combinations.
In terms of complexity, this is O($k^n$).
If a model has $n$ distinct algorithmic steps, the complexity of their ordering is O($n!$).
% If we consider the variations of equations in a mathematical model, and programming constructs in an algorithm, the level of complexity is similar or higher.
Model-developmental visualization is a great challenge in the field of visualization.
\end{itemize}

Hence the levels correspond to the questions to be asked and the complexity of the space of optional answers.
For example, given a financial prediction model, if an analyst uses visualization to demonstrate its effectiveness to an audience, it falls into workflow class $W_3$.
This is level 1 visualization, as the analyst knows or assumes the model to be correct.

If the analyst sequentially observes a financial data stream and some basic statistics about the data in order to capture some events, it more or less follows the same workflow $W_2$, and it is level 2 visualization.

If the analyst applies a prediction model to the input data streams, and then uses visualization to observe the input data and its basic statistics, to receive the predictions and recommendations computed by a machine process, and to reason about potential errors, such a process is encapsulated by $W_4$.
The analysis of errors and noise typically involves examination of the relationships among different events in the input data streams, statistical indicators, computed trends and recommendations.
It is more complex than observing events in a data stream sequentially.
This is level 3 visualization. 

If the analyst identifies that a prediction model does not perform satisfactorily, and attempts to optimize it by, for example, experimenting with various parameters in the model, this falls into workflow class $W_5$.
Alternatively, the analyst may wish to create a new prediction model based on a different economic theory, this falls into workflow class $W_6$.
When visualization is used to assist the analyst in exploring the parameter space or the model space, this is level 4 visualization.

% ===========
\begin{figure*}[t!]
\centering
\includegraphics[width=150mm]{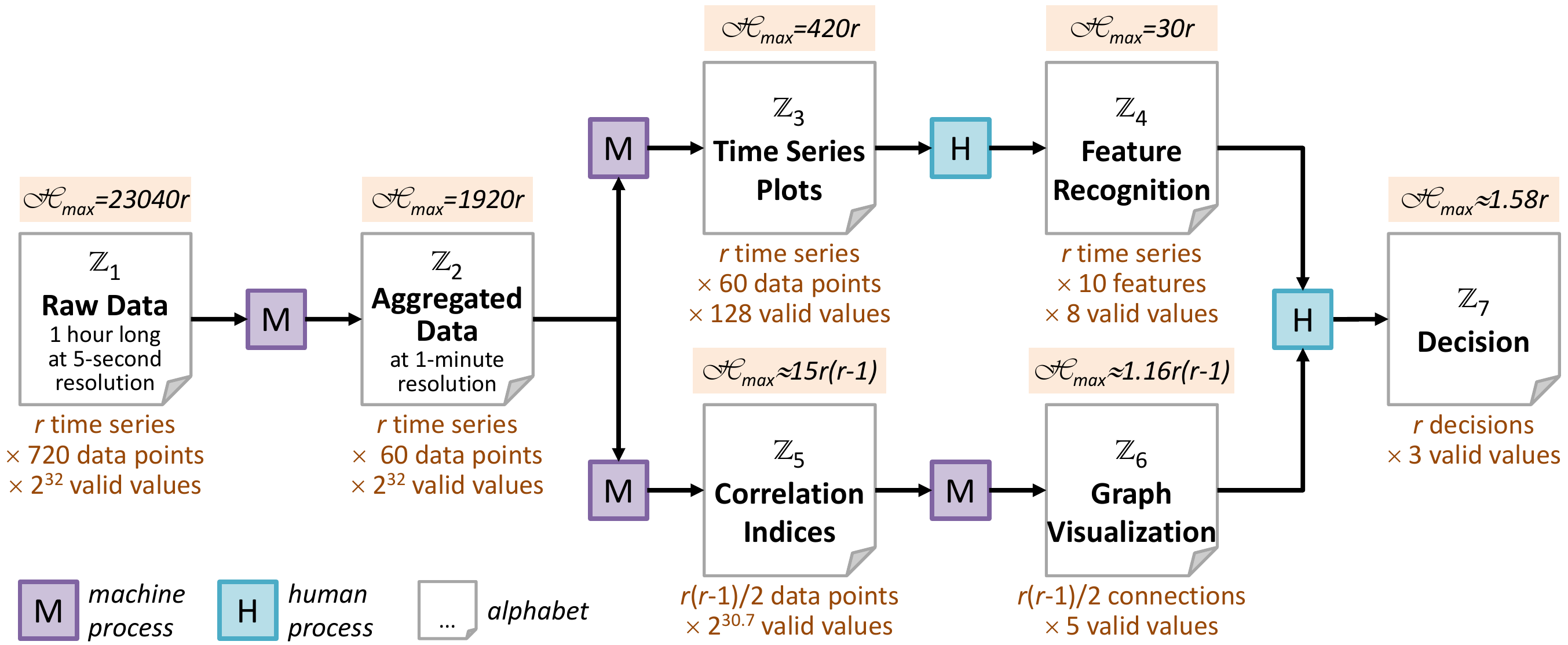} \\
\caption{An example transformation of alphabets during a data analysis and visualization process. From left to right, the initial alphabet corresponds to $r$ time series each capturing a share price at 5 second interval within an hour. For each time series, the 12 data points in every minute are then aggregated into a mean value. The $r$ time series is then visualized as line plots. The analyst identifies various features during the visualization, such as different levels of rise or fall, different speed, etc. Meanwhile, the analyst computes the correlation indices between each pair of time series and visualize these using, for instance, a circular graph plot, where correlation indices are mapped to five different colors. The analyst finally makes a decision for each of the $r$ shares as to buy, sell or hold. The maximal entropy $\mathcal{H}_{MAX}$ shows a decreasing trend from left to right.}
\label{fig:Alphabets}
\vspace{-2mm}
\end{figure*}

\section{An Information-Theoretic Abstraction}
\label{sec:Abstraction}

% ----------
\subsection{Alphabets and Letters}
The term \emph{data object} is an encompassing generalization of datum, data point, data sample, data record and dataset.
It contains a finite collection of quantitative and/or qualitative measures that are values of a finite set of variables.
For example, consider a univariate variable $X$ for recording the population of a country.
A value representing the UK population in 2010 is a datum, and thus a data object.
A collection of the individual population figures of $\mathsc{n}$ countries in 2010 is also a data object, where the $\mathsc{n}$ values may be considered as a sample of data points of $X$, or separate records of $\mathsc{n}$ variables $X_i\ (i=1,2,\ldots,\mathsc{n})$.
Similarly, a time series recoding the UK annual population between 1900 and 2010 is a data object.
The 111 values in the time series may be considered as data points of the same univariate variable $X$, or a multivariate record for time-specific variables $X_t\ (t=1900,1901,\ldots,2010)$.
Of course, the term data object can also refer to a multivariate data point that consists of values representing conceptually-different variables, such as the area, population and GDP of a country.

The generalization also encompasses datasets that are often regarded as ``unstructured''.
For example, a piece of text may be treated as a multivariate record of $\mathsc{m}$ characters, each of which is a value of a variable $C_j$ for encoding a letter, digit or punctuation mark at a specific position $j\ (j=1,2,\ldots,\mathsc{m})$ within the text.
Hence, the multivariate record is a data object.
Alternatively, we can consider a composite variable, $Y$, which encodes all possible variations of texts with $\mathsc{m}$ or fewer characters.
A specific text with $1 \leq k \leq \mathsc{m}$ characters is thus a value of $Y$.
This example also illustrates the equivalence between encoding a data object as a multivariate data record or encoding it as an instance of single composite variable. 

In this generalized context, let $Z$ be a variable, and $\mathbb{Z} = \{z_1, z_2, \ldots, z_\mathsc{m}\}$ be the set of all its valid values.
$Z$ may be a univariate, multivariate, or composite variable.
When $Z$ is a multivariate variable, each of its valid value, $z_i$, is a valid combination of valid values of individual univariate variables.
When $Z$ is a composite variable, we can flatten its hierarchy by encoding the hierarchical relationships explicitly using additional variables.
The flattened representation thus represents a multivariate variable.
Hereby $z_i$ is a valid combination of valid values of individual variables including the additional ones.
In information theory, such a set $\mathbb{Z}$ is referred to as an \emph{alphabet}, and each of its member $z_i$ as a \emph{letter}.

When the probability of every letter, $\mathbf{p}(z_i)$, is known or can be estimated, $\mathbf{p}$ is the \emph{probability mass function} for the set $\mathbb{Z}$.
Shannon introduced the measure of \emph{entropy}:
\vspace{-2mm}
\[
	\mathcal{H}(Z) = - \sum_1^\mathsc{m} \mathbf{p}(z_i) \log_2 \mathbf{p}(z_i)
\vspace{-2mm}
\]
\noindent for describing the \emph{level of uncertainty} of an alphabet. With the above $\log_2$-based formula, the unit of $\mathcal{H}(Z)$ is \emph{bit}.

% ----------
\subsection{Transformation of Alphabets}
In many data-intensive environments, the alphabet of raw input data may contain numerous letters.
For example, consider all valid time series of share prices within one hour period.
Assuming that the share price is updated every 5 seconds, there are 720 data points per time series.
Assuming that we represent share price at USD \$0.01 resolution using 32-bit unsigned integers, the minimum and maximum values are thus 0 and $2^{32}-1$ cents respectively.
(Note: Historically the highest share price in the US is 347,600 cents, i.e., $2^{18} < 347,600 < 2^{19}$).
If the probability of different time series were uniformly distributed, the entropy of this alphabet would be $23040 = 720 \times \log_2(2^{32})$ bits.
This is the \emph{maximal entropy} of this alphabet.
In practice, as many high values in the range $[0, 2^{32}-1]$ are very unlikely, and sudden changes between a very low value and a very high value (or vice versa) during a short period are also rare, the actual entropy is lower than $23040$ bits.

On the other hand, if we need to consider $r$ of such time series in order to make a decision, the size of the new alphabet will increase significantly.
Although some combinations of $r$ time series may be highly improbable, they may still be valid letters.
Hence the maximal entropy of this new alphabet is $23040r$ bits.
Let us consider such $r$ time series as the initial raw data for a data analysis and visualization process as illustrated in Fig. \ref{fig:Alphabets}.

One may find that the resolution of 1 data point per 5 seconds is not necessary, and choose to reduce it to 1 data point every minute by computing the average of 12 data points in each minute.
The average values may also be stored using 32-bit unsigned integers.
This aggregation results in a new alphabet, whose maximal entropy of $1920r = r \times 60 \times \log_2(2^{32})$ bits.
When we use line plots to visualize these $r$ time series, we may only be able to differentiate 128 data values per data point.
In this case, the maximal entropy is reduced to $r \times 60 \times \log_2(128) = 420r$ bits.

When one observes these $r$ time series, one may identify some specific features, such as [rise, fall, or flat], [slow, medium, or fast], [stable, uneven, or volatile] and so on.
These features become a new set of variables defined at the level of an hour-long time series.
If we construct a new alphabet based on these feature variables, its entropy will be much less than $23040r$ bits.
For example, if there are 10 feature variables and each with 8 valid values, the maximal entropy of this ``observational'' alphabet is $30r$ bits.

When one analyzes the relations among these $r$ time series, one may, for instance, compute the correlation indices between every pair of time series.
This yields $r(r-1)/2$ numbers.
Assuming that these are represented using 32-bit floating-point numbers, the maximal entropy of this ``analytical'' alphabet is around $15r(r-1)$ bits as the single precision floating-point format supports some $2^{30.7}$ values in $[-1, 1]$.
When we visualize these correlation indices by mapping them to, for instance, five colors representing $[-1, -0.5, 0, 0.5, 1]$, the entropy is reduced to $\log_2(5) r(r-1)/2 \approx 1.16 r (r-1)$ bits. 

One may wish to make a decision with three options, [buy, sell, or hold].
In this case, this ``decisional'' alphabet for each time series has only three letters.
The maximal entropy of this alphabet is less than 2 bits.
If a decision has to be made for all $r$ time series, we have less than $2r$ bits.
Fig. \ref{fig:Alphabets} illustrates the abovementioned changes of alphabets with different maximal entropy values.
The final alphabet ultimately defines the visualization task, while some intermediate alphabets may also capture subtasks in a data analysis and visualization process.

% ----------
\subsection{Measuring Cost-Benefit Ratio}
From Fig. \ref{fig:Alphabets}, one observation that we can make is that there is almost always a reduction of maximal entropy from the original data alphabet to the decisional alphabet.
This relates to one of the basic objectives in statistical inference, i.e., to optimize the process between the initial alphabet and the final alphabet with minimal loss of information that is ``important'' to the decision based on the final alphabet.
However, as visualization processes involve both machine-centric and human-centric mappings, it is necessary (i) to optimize both types of mapping in an integrated manner, (ii) to take into account ``soft'' information that can be introduced by human analysts during the process, (iii) to consider information loss as part of a cost-benefit analysis.

Let us consider a sequential workflow with $\mathsc{l}$ processing steps.
There are $\mathsc{l}+1$ alphabets along the workflow, 
Let $\mathbb{Z}_s$ and $\mathbb{Z}_{s+1}$ be two consecutive alphabets such that:
\[
	F_s: \mathbb{Z}_s \longrightarrow \mathbb{Z}_{s+1}
\]
\noindent where $F_s$ is a mapping function, which can be an analytical algorithm that extracts features from data, a visual mapping that transforms data to a visual representation, or a human decision process that selects an outcome from a set of options.

The cost of executing $F_s$ as part of a visualization process can be measured in many ways.
Perhaps the most generic cost measure is \emph{energy} since energy would be consumed by a computer to run an algorithm or to create a visualization, as well as by a human analyst to read data, view visualization, reason about a possible relationship, or make a decision.
We denote this generic measurement as a function $\mathcal{C}(F_s)$. 
While measuring energy usage by computers is becoming more practical \cite{Thiyagalingam:2013:CGF}, measuring that of human activities, especially cognitive activities may not be feasible in most situations.
A more convenient measurement is \emph{time}, $\mathcal{C}_{time}(F_s)$, which can be considered as an approximation of  $\mathcal{C}(F_s)$.
Another is a monetary measurement of computational costs or employment costs, which represent a subjective approximation from a business perspective.
Without loss of generality, we will use $\mathcal{C}(F_s)$ as our cost function in this section.

\vspace{2mm}
\noindent \textbf{\textsc{Definition 1} (Alphabet Compression Ratio).}
As shown in Fig. \ref{fig:Alphabets}, a mapping function (i.e., a machine or human processes) usually facilitates the reduction of data space at each stage of data processing though the reduction is not guaranteed. We can measure the level of reduction as the \emph{alphabet compression ratio} (ACR) of a mapping $F_s$:
\begin{equation}
\label{eq:CR}
	  \Psi_{ACR}(F_s) = \frac{\mathcal{H}(Z_{s+1})}{\mathcal{H}(Z_s)}
\end{equation}
\noindent where $\mathcal{H}$ is the Shannon entropy measure.
In a closed machine-centric processing system that meets the condition of a Markov chain, we have $\mathcal{H}(Z_{s}) \geq \mathcal{H}(Z_{s+1})$.
This is the \emph{data processing inequality} \cite{Cover:2006:book}. 
In such a system, $\Psi_{ACR}$ is a normalized and unitless entropy measure in $[0, 1]$ as first proposed by Golan in \cite{Golan:1988:PhD} (see also \cite{Golan:1996:book}).
However, Chen and J\"anicke pointed out that the Markov chain condition is broken in most visualization processes \cite{Chen:2010:TVCG}, and further examples were given in \cite{Chen:2013:S}.
Hence, we do not assume that $\mathcal{H}(Z_{s}) \geq \mathcal{H}(Z_{s+1})$ here since $F_s$ can be a human-centric transformation, unless one encodes all possible variants of ``soft'' information and knowledge in the initial data alphabet. 

%\vspace{2mm}
Meanwhile, given an output of an analytical process, $F_s$, an analyst will gain an impression about the input.
Considering the time series transformation in Fig. 2, for example, learning the mean price value for each minute, an analyst may have a conjecture about the 12 original data values.
Viewing a visualization of each time series plot in a resolution of 128 possible values per data point, an analyst may infer, estimate or guess the time series in its original resolution of $2^{32}$ possible values per data point.
Let us denote an impression about $Z_s$ as a variable $Z'_s$, which is a result of a mapping $G_s$ such that:

\[
	G_s: \mathbb{Z}_{s+1} \longrightarrow \mathbb{Z}'_s
\]

\noindent where $\mathbb{Z}'_s$ is the alphabet of this impression with a probability mass function representing the inferred or guessed probability of each letter in $\mathbb{Z}'_s$.
Note that $G_s$ is a reconstruction function, similar to what was discussed in \cite{Jaenicke:2011:CGF}.
In most cases, $G_s$ is only a rough approximation of the true inverse function $F^{-1}$.    
The difference between such an impression about $\mathbb{Z}'_s$ obtained from observing letters in $\mathbb{Z}_{s+1}$ and the actual $\mathbb{Z}_s$ is defined by Kullback-Leibler divergence (or relative entropy) \cite{Cover:2006:book}:
\[
	\mathcal{D}_{KL}(Z'_s || Z_s) = \mathcal{D}_{KL}(G(Z_{s+1}) || Z_s)\\
	= \sum_j \mathbf{p}(z'_{s,j}) \log_2 \frac{\mathbf{p}(z'_{s,j})}{\mathbf{q}(z_{s,j})}  
\]
\noindent where $z'_{s,j} \in \mathbb{Z}'_s$, and $z_{s,j} \in \mathbb{Z}_s$, and $\mathbf{p}$ and $\mathbf{q}$ are two probability mass functions associated with $\mathbb{Z}'_s$ and $\mathbb{Z}_s$ respectively.
$\mathcal{D}_{KL} = 0$  if and only if $\mathbf{p} = \mathbf{q}$, and $\mathcal{D}_{KL} > 0$ otherwise.
Note that $\mathcal{D}_{KL}$ is not a metric as it is not symmetric.
The definition of $\mathcal{D}_{KL}$ is accompanied by a precondition that $\mathbf{q} = 0$ implies $\mathbf{p} = 0$.

\vspace{2mm}
\noindent \textbf{\textsc{Definition 2} (Potential Distortion Ratio).} 
With the $\log_2$ formula, $\mathcal{D}_{KL}$ is also measured in bits.
The higher the number of bits is, the further is the deviation of the impression $\mathbb{Z}'_s$ from $\mathbb{Z}_s$.
The \emph{potential distortion ratio} (PDR) of a mapping $F_s$ is thus:
\begin{equation}
\label{eq:DR}
	  \Psi_{PDR}(F_s) = \frac{\mathcal{D}_{KL}(Z'_s || Z_s)}{\mathcal{H}(Z_s)}
\end{equation}
% \noindent Again this is a normalized and unitless entropy measure in $[0, 1]$.

Both $\Psi_{ACR}(F_s)$ and $\Psi_{PDR}(F_s)$ are unitless.
They can be used to moderate the cost of executing $F_s$, i.e., $\mathcal{C}(F_s)$.
Since $\mathcal{H}(Z_{s+1})$ indicates the intrinsic uncertainty of the output alphabet and $\mathcal{D}_{KL}(Z'_s || Z_s)$ indicates the uncertainty caused by $F_s$, the sum of $\Psi_{ACR}(F_s)$ and $\Psi_{PDR}(F_s)$ indicates the level of combined uncertainty in relation to the original uncertainty associated with $\mathbb{Z}_s$.

\vspace{2mm}
\noindent \textbf{\textsc{Definition 3} (Effectual Compression Ratio).} The \emph{effectual compression ratio} (ECR) of a mapping $F_s$ from $\mathbb{Z}_s$ to $\mathbb{Z}_{s+1}$ is a measure of the ratio between the uncertainty before a transformation $F_s$ and that after:

\begin{equation}
\label{eq:ECR}
 	\Psi_{ECR}(F_s) = \frac{\mathcal{H}(Z_{s+1}) + \mathcal{D}_{KL}(Z'_s || Z_s)}{\mathcal{H}(Z_s)} \quad \text{for }\mathcal{H}(Z_s)>0
\end{equation}

\noindent When $\mathcal{H}(Z_s) = 0$, it means that variable $Z_s$ has only one probable value, and it is absolute certain.
Hence, the transformation of $F_s$ is unnecessary in the first place.
The measure of ECR encapsulates the tradeoff between ACR and PDR, since deceasing ACR (i.e., more compressed) often leads to an increase of PDR (i.e., harder to infer $\mathbb{Z}_s$), and vice versa.
However, this tradeoff is rarely a linear (negative) correlation.
Finding the most appropriate tradeoff is thus an optimization problem, which is to be further enriched when we incorporate below the cost $\mathcal{C}(F_s)$ as another balancing factor. 

\vspace{2mm}
\noindent \textbf{\textsc{Definition 4} (Benefit).} We can now define the \emph{benefit} of a mapping $F_s$ from $\mathbb{Z}_s$ to $\mathbb{Z}_{s+1}$ as:

\begin{equation}
\label{eq:Benefit}
	  \mathcal{B}(F_s) =  \mathcal{H}(Z_s) - \mathcal{H}(Z_{s+1}) - \mathcal{D}_{KL}(Z'_s || Z_s)
\end{equation}

\noindent The unit of this information-theoretic measure it \emph{bit}.
When $\mathcal{B}(F_s) = 0$, the transformation does not create any change in the informational structure captured by the entropy.
In otherwords, there is no informational difference between observing variable $Z_s$ and observing $Z_{s+1}$.
When $\mathcal{B}(F_s) < 0$, the transformation has introduced more uncertainty, which is undesirable.
When $\mathcal{B}(F_s) > 0$, the transformation has introduced positive benefit by reducing the uncertainty.
This definition can be related to Shannon's grouping property \cite{Cover:2006:book}.

\vspace{2mm}
\noindent \textbf{\textsc{Theorem} (Generalized Grouping Property).}
Let $X$ be a variable that is associated with an $\mathsc{n}$-letter alphabet $\mathbb{X}$ and a normalized $\mathsc{n}$-dimensional discrete distribution $\mathbf{p}(x), x \in \mathbb{X}$.
When we group letters in $\mathbb{X}$ in to $\mathsc{m}$ subsets, we derive a new variable $Y$ with an $\mathsc{m}$-letter alphabet $\mathbb{Y}$ and a normalized $\mathsc{m}$-dimensional discrete distribution $\mathbf{q}(y), y \in \mathbb{Y}$.

\begin{equation}
\label{eq:Grouping}
	  \mathcal{H}(X) =  \mathcal{H}(Y) + \sum_{k=1}^\mathsc{m} \mathbf{q}(y_k) \mathcal{H}_k
\end{equation}

\noindent where $\mathcal{H}_k$ is the entropy of the local distribution of the original letters within the $k^{th}$ subset of $\mathbb{X}$.
Comparing Eq.\,(\ref{eq:Benefit}) and Eq.\,(\ref{eq:Grouping}), we can see that the last term on the right in Eq.,(\ref{eq:Grouping}) is replaced with the Kullback-Leibler divergence term in Eq.\,(\ref{eq:Benefit}).
The equality in Eq.\,(\ref{eq:Grouping}) is replaced with a measure of difference in Eq.\,(\ref{eq:Benefit}).
This is because of the nature of data analysis and visualization.
After each transformation $F_s$, the analyst is likely to infer, estimate or guess the local distribution within each subset, when necessary, from the observation of $X$ in the context of Eq.\,(\ref{eq:Grouping}) or $Z_{s+1}$ in the context of Eq.\,(\ref{eq:Benefit}) in conjunction with some ``soft'' information and knowledge, as mentioned in Section 1.

\vspace{2mm}
\noindent \textbf{\textsc{Definition 5} (Incremental Cost-Benefit Ratio).} The \emph{incremental cost-benefit ratio} (Incremental CBR) of a mapping $F_s$ from $\mathbb{Z}_s$ to $\mathbb{Z}_{s+1}$ is thus defined as the ratio between benefit $\mathcal{B}(F_s)$ and cost $\mathcal{C}(F_s)$.

\begin{equation}
\label{eq:ICBR}
	 \Upsilon(F_s) =  \frac{\mathcal{B}(F_s)}{\mathcal{C}(F_s)}
	= \frac{\mathcal{H}(Z_s) - \mathcal{H}(Z_{s+1}) - \mathcal{D}_{KL}(Z'_s || Z_s)}{\mathcal{C}(F_s)} 
\end{equation}

\noindent Note that we used cost as the denominator because (i) the benefit can be zero, while the cost of transformation cannot be zero as long as there is an action of transformation; (ii) it is better to associate a larger value to the meaning of more cost-beneficial. 

% \begin{equation}
% \label{eq:CE}
%	  \Upsilon(F) =   \mathcal{C}(F_s) \biggl(  \Psi_{CR} + \Psi_{PDR}  \biggr) = \mathcal{C}(F_s) \frac{\mathcal{H}(Z_{s+1}) + \mathcal{D}_{KL}(Z'_s || Z_s)} {\mathcal{H}(Z_s)}
% \end{equation}

Given a set of cascading mapping functions, $F_1, F_2, \ldots, F_\mathsc{l}$, which transform alphabets from $\mathbb{Z}_1$ to $\mathbb{Z}_{\mathsc{l}+1}$, we can simply add up their costs and benefits as:
\begin{align*}
	\mathcal{C}_{total} &= \sum_{s=1}^\mathsc{L} \mathcal{C}(F_s)\\
	\mathcal{B}_{total} &= \sum_{s=1}^\mathsc{L} \mathcal{E}(F_s)
	= \mathcal{H}(Z_1) - \mathcal{H}(Z_{\mathsc{l}+1}) - \sum_{s=1}^\mathsc{L} \mathcal{D}_{KL}(Z'_s || Z_s)
\end{align*}

\noindent The \emph{overall cost-benefit ratio} (Overall CBR) is thus $\mathcal{B}_{total}/\mathcal{C}_{total}$.

For workflows containing parallel mappings, the merge of CBR at a joint partly depends on the semantics of the cost and benefit measures.
If we are concerned about the energy, or monetary cost, the simple summation of cost measures arrived at a joint makes sense.
If we are concerned about the time taken, we may compute the maximum cost at a joint.
If all parallel branches arriving at a joint contain only machine-centric processes, the benefit is capped by the entropy at the beginning of the branching-out.
The combined benefit can be estimated by taking into account the mutual information between the arriving alphabets.
When these parallel branches involve human-centric processing, ``soft'' information will be added into the process.
The combined benefit can be estimated in the range between the maximum and the summation of the arriving benefit measures.

In this paper, we largely focus on the workflows for conducting data analysis and visualization.
Our formulation of cost-benefit analysis can be extended to include the cost of development and maintenance.
It is more appropriate to address such an extension in future work.

% ===========
\begin{figure}[t!]
\centering
\includegraphics[width=76mm]{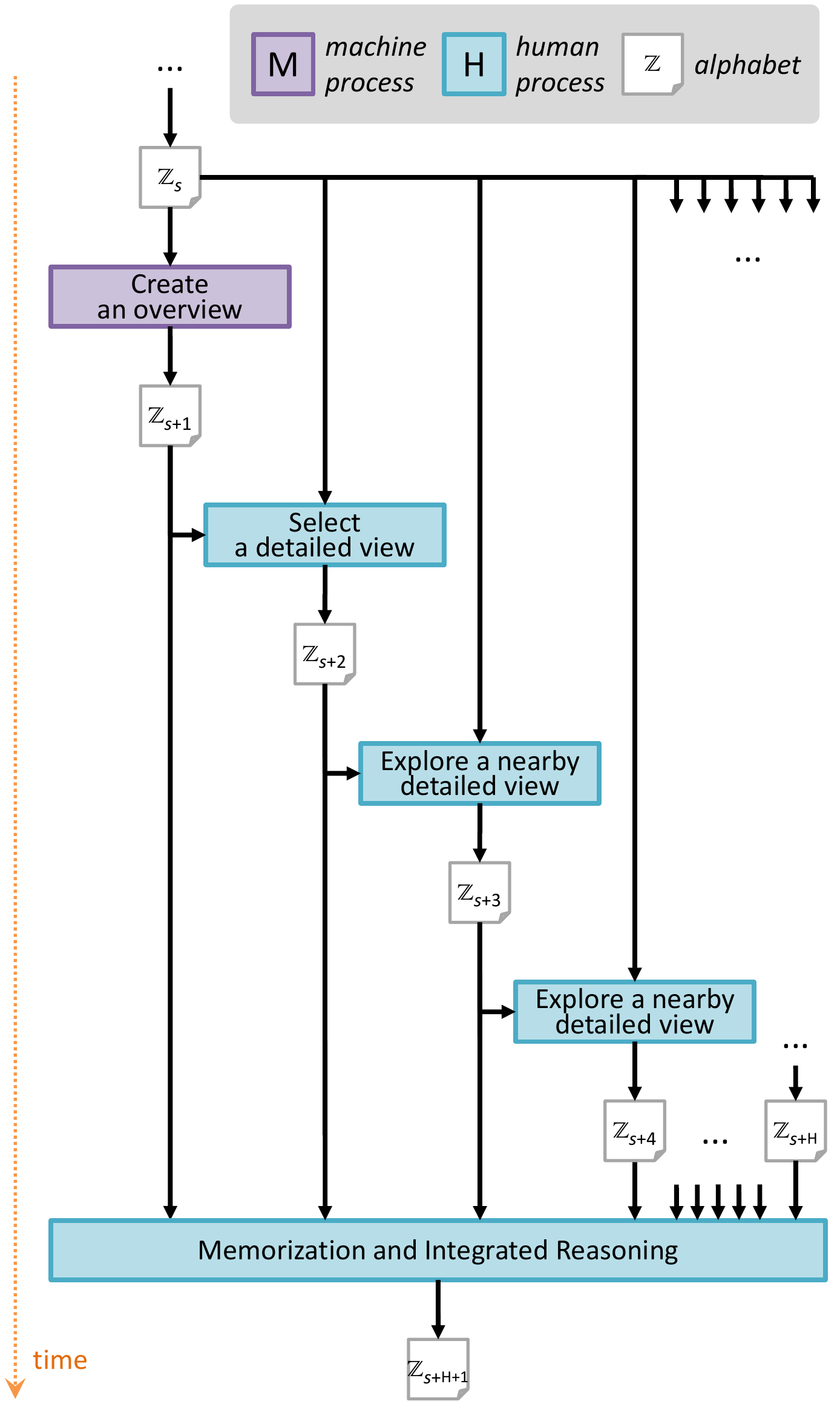} \\
\caption{Interaction is one of the means for introducing ``soft'' information into a visualization process. This figure shows a sequence of interactions for overview first and detailed on demand. At some stage of a visualization process, the system receives a large detailed visual representation $z \in \mathbb{Z}_s$. It creates an overview. A viewer selects a part of the overview and requests a detailed view, which is part of $z$. From this detailed view, the viewer explores a few nearby detailed views. At some stage, the viewer decides to finish the exploration and makes up his/her mind about something based on the overview and partial observation of the detailed visual representation.
At each step, all possible valid inputs and outputs of a transformation are letters of an alphabet.}  
\label{fig:Overview}
\vspace{-2mm}
\end{figure}

\section{Examples of Workflow Analysis}
\label{sec:Demonstration}

In this section, we consider several successful visualization processes in the literature.
We analyze their cost-benefit ratios in comparison with possible alternative processes.
The comparison serves as initial validation of the information-theoretic measures proposed in the previous section.
Like most theoretic development, the validation of the proposed information-theoretic measures should be, and is expected to be, a long-term undertake, along with the advancement of techniques and the increasing effort for collecting performance data about various human-centric processes, e.g., through empirical studies.   

% ----------
\subsection{Interaction in Visualization}
In data analysis and visualization, human-computer interaction plays a significant role in breaking the condition of the \emph{data processing inequality} \cite{Chen:2010:TVCG}.
It enables human analysts to introduce ``soft'' information and knowledge into such a process.
Here we consider that the initial alphabet at the beginning of the process represents ``hard'' data, e.g., $\mathbb{Z}_1$ for representing variants of $r$ time series.
The ``soft'' information and knowledge is ``external'' to the process, which can no longer be a closed system.

Interaction has been studied extensively in the context of visualization (e.g.,
\cite{Chuah:1996:InfoVis,Tweedie:1997:CHI,Pfitzner:2003:APSIV,Ward:2004:EuroVis,Yi:2007:TVCG}).
One of the commonly-used forms of interaction is ``overview first, zoom and detailed on demand'' \cite{Shneiderman:1996:VL}.
It may feature several types of actions, including \emph{select}, \emph{explore}, \emph{abstract/elaborate} and \emph{filter} as defined by Yi \emph{et al.} \cite{Yi:2007:TVCG}.
Fig. \ref{fig:Overview} illustrates such a process with the abstract notion presented in Section \ref{sec:Abstraction}.
Under an information-theoretic framework, the input and output of each transformation are considered in a holistic manner, i.e., as alphabets (e.g., all variants of images that may be displayed in a context) rather than individual letters (e.g., an image).

One may imagine a very large image (or map) as an instance of alphabet $\mathbb{Z}_s$ at the top of Fig. \ref{fig:Overview}.
An interactive system first presents viewers with an overview, which is an instance of alphabet $\mathbb{Z}_{s+1}$.
A viewer may select a part of the overview and apply a zoom-in operation.
The detailed view at that location is an instance of alphabet $\mathbb{Z}_{s+2}$, which is a subset of $\mathbb{Z}_{s}$.
From this detailed view, the viewer may choose to explore to a nearby location, and so on.
At some stage, the viewer decides to finish the exploration and makes up his/her mind about something based on the overview and parts of the full image (or map) representation that has been explored so far.
The examples of ``soft'' information and knowledge in this case may include: how important the individual subsets of $\mathbb{Z}_s$ are to the viewer, and which direction of exploration from one subset to the next is more promising.

Let us consider the incremental CBR (cost-benefit ratio) of the overview transformation.
Different techniques can be used to compute overview visualization, yielding different ACR (alphabet compression ratio) and PDR (potential distortion ratio).
As this is a machine-centric process, the term $\mathcal{D}_{KL}(Z'_s || Z_s)$ in PDR can be replaced with
\[
	\mathcal{H}(Z_s) - \mathcal{I}(Z_s, Z_{s+1})
\]
\noindent where $\mathcal{I}$ is the mutual information between alphabets $\mathbb{Z}_s$ and $\mathbb{Z}_{s+1}$ with an assumption that a prefect inverse mapping $F_s^{-1}$ from $\mathbb{Z}_{s+1}$ to $\mathbb{Z}_s$ can infer all mutual information, but no more than that.
Hence, we can rewrite Eq.\,(\ref{eq:ICBR}) as:

\begin{equation}
\label{eq:ICBR-M}
	 \Upsilon(F_s) =  \frac{\mathcal{B}(F_s)}{\mathcal{C}(F_s)}
	= \frac{\mathcal{I}(Z_s, Z_{s+1}) - \mathcal{H}(Z_{s+1})}{\mathcal{C}(F_s)} 
\end{equation}

Recall an example discussed in \cite{Chen:2010:TVCG}, where two different overview techniques for flow visualization were used to illustrate the optimization based on mutual information $\mathcal{I}$.
This criterion is consistent with Eq.\,(\ref{eq:ICBR}) when one assumes that that the two techniques maintain the same entropy for the output alphabet $\mathbb{Z}_{s+1}$, while incurring the same cost $\mathcal{C}(F_s)$.
Eq.\,(\ref{eq:ICBR-M}) is thus an extension of what proposed in \cite{Chen:2010:TVCG}.

In Fig. \ref{fig:Overview}, the transformations following $\mathbb{Z}_{s+1}$ are all human-centric processes.
We can observe that these transformations will incur costs such as cognitive effort and time for interaction.
One important consideration is the \emph{prior knowledge} about $\mathbb{Z}_s$, i.e., how much information is already known to the viewers, and how much is uncertain.
Considering the same flow visualization example as in \cite{Chen:2010:TVCG}, one may estimate:
\begin{itemize}
\vspace{-2mm}
\item How likely do viewers know that $\mathbb{Z}_s$ are texture-based representations of vector fields?
\vspace{-2mm}
\item How confident are viewers about the correctness of the feature-extraction technique used to create an overview?
\vspace{-2mm}
\item How long have the viewers been working on the simulation model that generates the vector fields being visualized?
\end{itemize} 

Such inference can be translate to an estimation about the term $\mathcal{D}_{KL}(Z'_s || Z_s)$ in Eqs.\,\ref{eq:ECR} and \ref{eq:ICBR}.
Some further optimization can be often implemented on top of the basic form of \emph{overview first and details on demand}.
In many scenarios, we often observe that an experienced viewer may find step-by-step zoom operations frustrating, as the viewer knows exactly where is the interesting part of a detailed representation.
For example, in flow simulation, scientists often work on the same simulation problem for months, and have a good mental overview about $\mathbb{Z}_s$.
In such a case, when the interactive visualization system has a fast track for reaching a specific detailed view (e.g., the last location visited), it reduces the cost of step-by-step zoom operations.
However, this approach may not be applicable to an online map system, where each search session is likely for a new search task.
 
% -------
\begin{figure*}[t!]
\centering
\includegraphics[width=145mm]{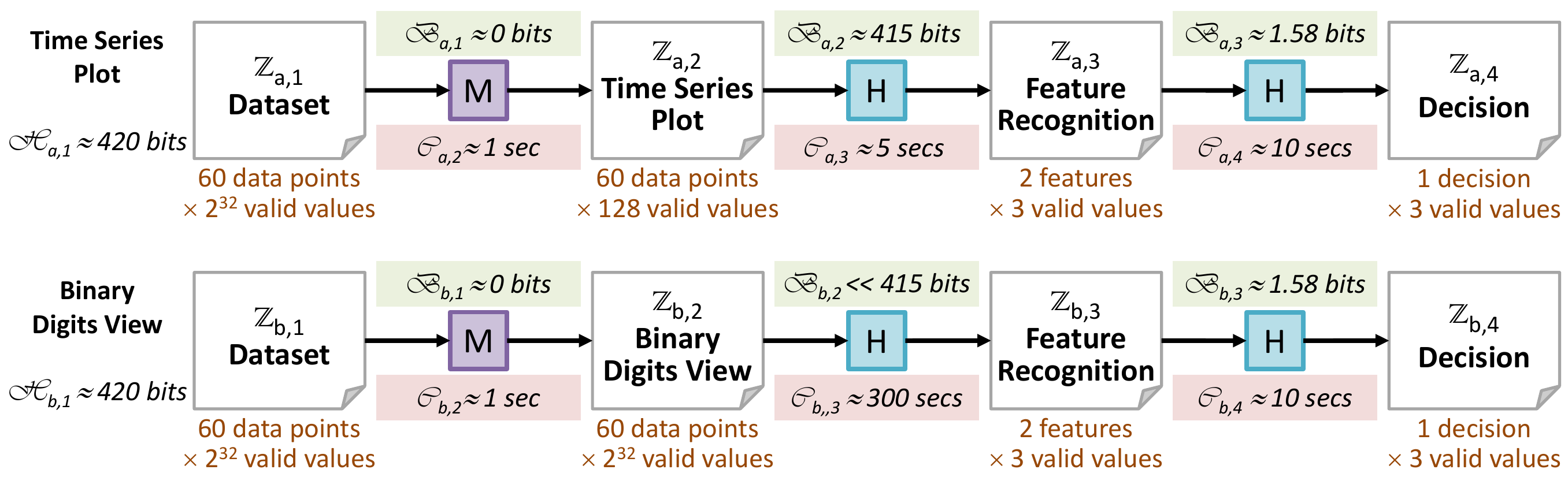} \\
\caption{Comparison between time series plot and binary digitals view for disseminative visualization. The same legend in Fig. \ref{fig:Alphabets} applies. The estimated benefit and cost values here are based on heuristic reasoning and for an illustrative purpose. For example, for $\mathcal{B}_{a,2}$, we consider two feature variables [stable, uneven, volatile] and [rise, fall, flat]. Hence the maximal entropy of $\mathbb{Z}_{a,3}$ is about $3.17$ bits.
As the $\mathcal{D}_{KL}$ term for $\mathcal{B}_{a,2}$ will indicate some uncertainty, the estimated benefit is $420-3.17-\mathcal{D}_{KL}(\mathbb{Z}'_{a,2}||\mathbb{Z}_{a,2}) \approx 415$ bits. Meanwhile, $\mathcal{D}_{KL}$ for $\mathcal{B}_{b,2}$ is much higher.
}
\label{fig:DV}
\vspace{-2mm}
\end{figure*}

\subsection{Disseminative Visualization}
\label{sec:DV}
The history of time series plot can be traced back more than a millennium ago.
If success is measured by usage, it is no doubt one of the most successful visual representations.
However, its display space utilization is rather poor in comparison with a binary digits view \cite{Chen:2010:TVCG}.
Fig. \ref{fig:DV} shows two such representations that are used as disseminative visualization for a scenario in Fig. \ref{fig:Alphabets}.
The dataset being displayed is a time series with 60 data points, i.e., an instance of $\mathbb{Z}_2$ in Fig. \ref{fig:Alphabets}.
Assume that the value of this particular share has been largely moving between 100 and 200 cents.
Hence the entropy of $\mathbb{Z}_{a,1} = \mathbb{Z}_{b,1}$ is estimated to be about 420 bits, significantly below the maximal entropy of the data representation.

The binary digits view uses a $2\!\times\!2$ pixel-block per digit, and requires $32\!\times\!60$ blocks (7,680 pixels) for the plotting canvas.
Using the same number of pixels, $128\!\times\!60$, the time series plot is an instance of $\mathbb{Z}_{a,2}$.
During dissemination, the presenter (or analyst) points out ``stable'' and ``rise'' features to a viewer (or client), suggesting a decision ``to hold''.
The overall CBRs for the two pipelines in Fig. \ref{fig:DV} are:

\begin{equation}
\label{eq:DV-plot}
	\Upsilon_{plot} = \sum_{j=1}^3 \frac{\mathcal{H}(Z_{a,j+1}) + \mathcal{D}_{KL}(Z'_{a,j} || Z_{a,j})}{\mathcal{C}(F_{a,j})}
\end{equation}
 
\begin{equation}
\label{eq:DV-binary}
	\Upsilon_{binary} = \sum_{j=1}^3 \frac{\mathcal{H}(Z_{b,j+1}) + \mathcal{D}_{KL}(Z'_{b,j} || Z_{b,j})}{\mathcal{C}(F_{b,j})}
\end{equation}

To the presenter, the decision ``to hold'' has already been made, and the total CBR would be zero for either workflow.
For a viewer unfamiliar with binary representations, the binary digits view is almost undecipherable.
Even for a pair of untrained eyes, recognizing features such as ``stable'' and ``rise'' would take a while.
The inverse mapping from the features pointed out by the presenter is also rather uncertain, hence a high value for the $\mathcal{D}_{KL}$ term in $\mathcal{B}_{b,2}$.   
The binary digits view thereby incurs a huge cost at the feature recognition step, while bringing lower benefit.
This mathematically explains the merits of time series plot over a spatially-compact binary digits view. 

% -------
\subsection{Observational Visualization}
\label{sec:OV}
The example in Section \ref{sec:DV} can also be considered in the context of observational visualization, where an analyst creates visualization for him/herself.
Similar abstract reasoning and step-by-step inference can be carried out, just as in the previous example, likely for a much larger input data alphabet (e.g., with $r$ time series and $t$ hours). 
 
Let us consider a different example of observational visualization.
Legg \emph{et al.} reported an application of visualization in sports \cite{Legg:2012:CGF}.
The Welsh Rugby Union required a visualization system for in-match and post-match analysis.
One of the visualization tasks was to summarize events in a match, facilitating external memorization.
The input datasets are typically in the form of videos including data streams during a match, and can be generalized to include direct viewing of a match in real-time.
The alphabet is thus huge.
The objective for supporting external memorization is to avoid watching the same videos repeatedly.
Especially during a half-time interval, coaches and players cannot afford much time to watch videos.

The workflow can be coarsely divided into three major transformations, namely $F_a$: transforming real-world visual data to events data, $F_b$: transforming events data to visualization, and $F_c$: transforming observations to judgments and decisions.
Clearly, transformation $F_c$ should be performed by coaches and other experts.
For transformation $F_a$, two options were considered: $F_{a,1}$ for computers to detect events, and $F_{a,2}$ for humans to detect events.
For transformation $F_b$, two options were considered: $F_{b,1}$ statistical graphics, and $F_{b,2}$ glyph-based event visualization.
For $F_{a,1}$ and $F_{a,2}$, the letters of the output alphabet are multivariate data objects describing what type of event, when and where it happens, and who are involved.
This alphabet is much smaller than the input alphabet for real-world visual data.

The team did not find any suitable computer vision techniques that could be used to detect events and generate the corresponding data objects in this application.
The accuracy of available techniques were too low, hence the $\mathcal{D}_{KL}$ term for $F_{a,1}$ will yield a high-level of uncertainty.
Using a video annotation system, an experienced sports analyst can generate more accurate event data during or after a match.
For an 80 minute Rugby match, the number of data objects generated is usually in hundreds and sometimes in thousands.
Hence statistics can be obtained, and then visualized using statistical graphics.
However, it is difficult for coaches to make decisions based on statistical graphics, as it is difficult to connect statistics with episodic memory about events.
Such a difficulty corresponds to a high-level of uncertainty resulting from the $\mathcal{D}_{KL}$ term for $F_{b,1}$.
On the other hand, the direct depiction of events using glyphs can stimulate episodic memory much better, yielding a much lower-level uncertainty in the $\mathcal{D}_{KL}$ term for $F_{b,2}$.
The team implemented $F_{a,2}$ and $F_{b,2}$ transformations as reported in \cite{Legg:2012:CGF}, while $F_{b,1}$ was also available for other tasks.

\subsection{Analytical Visualization}
\label{sec:AV}
Oelke \emph{et al.} studied a text analysis problem using visual analytics \cite{Oelke:2010:VAST}.
They considered a range of machine-centric and human-centric transformations in evaluating document readability.
For example, the former includes 141 text feature variables, and their combinations.
The latter includes four representations at three different levels of details.
Since different combinations of machine-centric and human-centric transformations correspond to different visual analytics pipelines,
their work can be seen as an optimization effort.
Through experimentation and analysis, they confirmed the need for enabling analysts to observe details at the sentence or block levels.
Over-aggregation (e.g., assigning a readability score to each document) is not cost beneficial, as the tradeoff between the alphabet compression ratio (ACR) and the potential distortion ratio (PDR) is in favor of PDR.  

% ----------
\subsection{Model-developmental Visualization}
\label{sec:MV}
In \cite{Tam:2011:CGF}, Tam \emph{et al.} compared a visualization technique and a machine learning technique in generating a decision tree as a model for expression classification.
The input to this model development exercise is a set of annotated videos, each of which records one of four expressions [anger, surprise, sadness, smile].
The output is a decision tree that is to be used to classify videos automatically with reasonable accuracy.
It is thus a \emph{data analysis and visualization process} for creating a \emph{data analysis model}. 
Although this sounds as a conundrum, it fits well within the scope of visualization.    
Tam \emph{et al.} approached this problem through a series of transformations.
The first transformation $F_a$ identifies 14 different facial features in each video, and records it temporal changes using a geometric or texture measurement.
This results in 14 different alphabets of time series.
The second transformation $F_b$ characterizes each time series using 23 different parameters.
This results in a total of $322 = 14 \times 23$ variables.
At the end of the second transformation, each video becomes a 322-variate data object.

For the visualization-based pipeline, the third transformation $F_{c,1}$ generates a parallel coordinate plot with 322 axes.
This is followed by the fourth transformation $F_{d,1}$, where two researchers laid the big plot on the floor and spent a few hours to select the appropriate variables for constructing a decision tree.
For the machine-learning based pipeline, the team used a public-domain tool, C4.5, as the third transformation $F_{c,2}$, which generates a decision tree from a multivariate dataset automatically.

In terms of time cost, transformation  $F_{c,2}$ took much less time than transformations  $F_{c,1}$  and $F_{d,1}$ together.
In terms of performance, the decision tree created by  $F_{c,1}$  and $F_{d,1}$ was found slightly more accurate than that resulting from  $F_{c,2}$.
From further analysis, they learned that
(i) handling real values has been a challenge in automatic generation of decision trees;
(ii) the two researchers did not rely solely on the parallel coordinates plot to choose variables, their ``soft'' knowledge about the underlying techniques used in transformations $F_a$ and $F_b$ also contributed to the selection.
Such ``soft'' knowledge reduces the uncertainty expressed by the $\mathcal{D}_{KL}$ term in Eq.\,\ref{eq:Benefit}.
This example demonstrates the important role of visualization in model development. 

\vspace{-1mm}
% ===========
\section{Conclusions}
\label{sec:Conclusions}
In this paper, we have proposed an information-theoretic measure for offering a mathematical explanation as to what may have been optimized in successful visualization processes.
We have used several examples in the literature to demonstrate its explanatory capability for both machine-centric and human-centric transformations in data analysis and visualization.
One question that naturally occurs is how one may use such a theoretical measure in a practical environment.
We consider this question in three stages.

(i) At present, it is important for us to recognize that the overall objective of data analysis and visualization corresponds to the reduction of Shannon entropy from the original data alphabet to the decisional alphabet. There is a cost associated with this reduction process.
It is also necessary to recognize that the benefit of such reduction at each incremental step is likely to be weakened by the uncertainty of an approximated inverse mapping, i.e., the $\mathcal{D}_{KL}$ term in Eq.\,\ref{eq:Benefit}. This uncertainty can be caused by inaccuracy or aggressive aggregation of a machine-centric transformation, as well as by human factors such as visual uncertainty \cite{Dasgupta:2012:CGF} and lack of understanding and experience.

(ii) Next, we can learn from cost-benefit analysis in social sciences, where quantitative and qualitative methods are integrated together to optimize various business and governmental processes in a systematized manner.
Once a visualization process is defined as a transformation-based pipeline, we can estimate the cost for each transformation.
We should start to define alphabets and estimate the uncertainty measures associated with them.  

(iii) Historically, theoretical advancements were often part of long-term co-evolution with techniques and processes for measurements. This suggests that in the future we will be able to optimize visualization processes in a more quantitative manner.
It also suggests that in visualization, empirical studies are not only for evaluating hypotheses but also for collecting measurements that can potentially be used in process optimization.
  
\vspace{-1mm}
\section*{Acknowledgment}
Both authors are members of \emph{Info-Metrics Institute} and would like to thank the Institute for providing a stimulating environment and financial support for this interdisciplinary research.

\bibliographystyle{abbrv}
%%use following if all content of bibtex file should be shown
%\nocite{*}
\bibliography{ChenGolan2015}

\end{document}